\documentclass{mn2e}  
\usepackage{epsf} 
\usepackage{amsmath} 
\begin{document}

\newcommand{\getinptfile}
{\typein[\inptfile]{Input file name}
\input{\inptfile}}

\newcommand{\mysummary}[2]{\noi {\bf SUMMARY}#1 \\ \noi \sl #2 \\ \capline 
	\hspace{-.13in} \raisebox{.0in}{$\sqcap$} \rm }  
\newcommand{\mycaption}[2]{\caption[#1]{\footnotesize #2}} 
\newcommand{\capline}{\mbox{}\hrulefill}
\newcommand{\mysection}[2]{ 
\section{\uppercase{\normalsize{\bf #1}}} \def\junksec{{#2}} } %
\newcommand{\mychapter}[2]{ \chapter{#1} \def\junkchap{{#2}}  
\def\thesection{\arabic{chapter}.\arabic{section}}
\def\thesubsection{\thesection.\arabic{subsection}}
\def\thesubsubsection{\thesubsection.\arabic{subsubsection}}
\def\theequation{\arabic{chapter}.\arabic{equation}}
\def\thefigure{\arabic{chapter}.\arabic{figure}}
\def\thetable{\arabic{chapter}.\arabic{table}}
}
\newcommand{\mysubsection}[2]{ \subsection{#1} \def\junksubsec{{#2}} }
\def\thenote{\addtocounter{footnote}{1}$^{\scriptstyle{\arabic{footnote}}}$ }

\newcommand{\myfm}[1]{\mbox{$#1$}}
\def\spose#1{\hbox to 0pt{#1\hss}}	
\def\ltabout{\mathrel{\spose{\lower 3pt\hbox{$\mathchar"218$}} 
     \raise 2.0pt\hbox{$\mathchar"13C$}}}
\def\gtabout{\mathrel{\spose{\lower 3pt\hbox{$\mathchar"218$}}
     \raise 2.0pt\hbox{$\mathchar"13E$}}}
\newcommand{\ltsim}{\raisebox{-0.5ex}{$\;\stackrel{<}{\scriptstyle \backslash}\;$}}
\newcommand{\simlt}{\ltsim}
\newcommand{\simgt}{\gtsim}
%
\newcommand{\unit}[1]{\ifmmode \:\mbox{\rm #1}\else \mbox{#1}\fi}
\newcommand{\ze}{\ifmmode \mbox{z=0}\else \mbox{$z=0$ }\fi }

%
\newcommand{\boldv}[1]{\ifmmode \mbox{\boldmath $ #1$} \else 
 \mbox{\boldmath $#1$} \fi}
%
\renewcommand{\sb}[1]{_{\rm #1}}%
\newcommand{\expec}[1]{\myfm{\left\langle #1 \right\rangle}}
\newcommand{\mone}{\myfm{^{-1}}}
\newcommand{\half}{\myfm{\frac{1}{2}}}
\newcommand{\nth}[1]{\myfm{#1^{\small th}}}
\newcommand{\ten}[1]{\myfm{\times 10^{#1}}}
\newcommand{\abs}[1]{\mid\!\! #1 \!\!\mid}
\newcommand{\as}{a_{\ast}}
\newcommand{\asr}{(a_{\ast}^{2}-R_{\ast}^{2})}
\newcommand{\bvm}{\bv{m}}
\newcommand{\calf}{{\cal F}}
\newcommand{\calI}{{\cal I}}
\newcommand{\calm}{{v/c}}
\newcommand{\calminf}{{(v/c)_{\infty}}}
\newcommand{\calQ}{{\cal Q}}
\newcommand{\calR}{{\cal R}}
\newcommand{\calw}{{\it W}}
\newcommand{\co}{c_{o}}
\newcommand{\cs}{C_{\sigma}}
\newcommand{\cst}{\tilde{C}_{\sigma}}
\newcommand{\cv}{C_{v}}
\def\dbar{{\mathchar '26\mkern-9mud}}	
\newcommand{\deldelr}{\frac{\partial}{\partial r}}
\newcommand{\deldelR}{\frac{\partial}{\partial R}}
\newcommand{\deldeltheta}{\frac{\partial}{\partial \theta} }
\newcommand{\deldelphi}{\frac{\partial}{\partial \phi} }
\newcommand{\ddotrc}{\ddot{R}_{c}}
\newcommand{\ddotxc}{\ddot{x}_{c}}
\newcommand{\dotrc}{\dot{R}_{c}}
\newcommand{\dotxc}{\dot{x}_{c}}
\newcommand{\Estar}{E_{\ast}}
\newcommand{\grpsi}{\Psi_{\ast}^{\prime}}
\newcommand{\kboltz}{k_{\beta}}
\newcommand{\levi}[1]{\epsilon_{#1}}
\newcommand{\limaso}[1]{$#1 ( a_{\ast}\rightarrow 0)\ $}
\newcommand{\limasinfty}[1]{$#1 ( a_{\ast}\rightarrow \infty)\ $}
\newcommand{\limrinfty}[1]{$#1 ( R\rightarrow \infty,t)\ $}
\newcommand{\limro}[1]{$#1 ( R\rightarrow 0,t)\ $}
\newcommand{\limrso}[1]{$#1 (R_{\ast}\rightarrow 0)\ $}
\newcommand{\limxo}[1]{$#1 ( x\rightarrow 0,t)\ $}
\newcommand{\limxso}[1]{$#1 (\xs\rightarrow 0)\ $}
\newcommand{\ls}{l_{\ast}}
\newcommand{\Ls}{L_{\ast}}
\newcommand{\mean}[1]{<#1>}
\newcommand{\ms}{m_{\ast}}
\newcommand{\Ms}{M_{\ast}}
\def\nb{{\sl N}-body }
\def\nbt{{\sf NBODY2} }
\def\nb1{{\sf NBODY1} }
\newcommand{\nuoned}{\nu\sb{1d}}
\newcommand{\ra}{\rightarrow}
\newcommand{\Ra}{\Rightarrow}
\newcommand{\rc}{r_{c} } 
\newcommand{\Rc}{R_{c} } 
\newcommand{\res}[1]{{\rm O}(#1)}
\newcommand{\rnsa}{(r^{2}-a^{2})}
\newcommand{\Rnsa}{(R^{2}-a^{2})}
\newcommand{\rs}{r_{\ast}}
\newcommand{\Rs}{R_{\ast}}
\newcommand{\Rsa}{(R_{\ast}^{2}-a_{\ast}^{2})}
\newcommand{\sa}{\sigma } 
\newcommand{\sac}{\sigma_{c} } 
\newcommand{\sas}{\sigma_{\ast} } 
\newcommand{\sasp}{\sigma^{\prime}_{\ast}}
\newcommand{\saxs}{\sigma_{\ast} } 
\newcommand{\sech}{{\rm sech}}
\newcommand{\tff}{t\sb{ff}} 
\newcommand{\ti}{\tilde}
\newcommand{\trel}{t\sb{rel}}
\newcommand{\ts}{\tilde{\sigma} } 
\newcommand{\tss}{\tilde{\sigma}_{\ast} } 
\newcommand{\vcol}{v\sb{col}}
\newcommand{\vs}{v_{\ast}  } 
\newcommand{\vsp}{v^{\prime}_{\ast}}
\newcommand{\vxs}{v_{\ast}  } 
\newcommand{\xs}{x_{\ast}}
\newcommand{\xc}{x_{c} } 
\newcommand{\xistar}{\xi_{\ast}}
\newcommand{\rmd}{\ifmmode \:\mbox{{\rm d}}\else \mbox{ d}\fi }
\newcommand{\rmD}{\ifmmode \:\mbox{{\rm D}}\else \mbox{ D}\fi }
\newcommand{\valfven}{v_{{\rm Alfv\acute{e}n}}}

%
\newcommand{\noi}{\noindent}
\newcommand{\bc}{boundary condition }
\newcommand{\bcs}{boundary conditions }
\newcommand{\Bcs}{Boundary conditions }
\newcommand{\lhs}{left-hand side }
\newcommand{\rhs}{right-hand side }
\newcommand{\wrt}{with respect to }
\newcommand{\iras}{{\sl IRAS }}
\newcommand{\cobe}{{\sl COBE }}
\newcommand{\Oh}{\myfm{\Omega h}}
%
\newcommand{\etal}{{\em et al.\/ }}
\newcommand{\eg}{{\em e.g.\/ }}
\newcommand{\etc}{{\em etc.\/ }}
\newcommand{\ie}{{\em i.e.\/ }}
\newcommand{\viz}{{\em viz.\/ }}
\newcommand{\cf}{{\em cf.\/ }}
\newcommand{\via}{{\em via\/ }}
\newcommand{\apriori}{{\em a priori\/ }}
\newcommand{\adhoc}{{\em ad hoc\/ }}
\newcommand{\viceversa}{{\em vice versa\/ }}
\newcommand{\versus}{{\em versus\/ }}
\newcommand{\qed}{{\em q.e.d. \/}}
\newcommand{\<}{\thinspace}
%
\newcommand{\km}{\unit{km}}
\newcommand{\kms}{\unit{km~s\mone}}
\newcommand{\kmsa}{\unit{km~s\mone~arcmin}}
\newcommand{\kpc}{\unit{kpc}}
\newcommand{\mpc}{\unit{Mpc}}
\newcommand{\hkpc}{\myfm{h\mone}\kpc}
\newcommand{\hmpc}{\myfm{h\mone}\mpc}
\newcommand{\parsec}{\unit{pc}}
\newcommand{\cm}{\unit{cm}}
\newcommand{\yr}{\unit{yr}}
\newcommand{\au}{\unit{A.U.}}
\newcommand{\AU}{\au}
\newcommand{\gm}{\unit{g}}
\newcommand{\solar}{\myfm{_\odot}}
\newcommand{\solarm}{\unit{M\olar}}
\newcommand{\Lsun}{\unit{L\solar}}
\newcommand{\Rsun}{\unit{R\solar}}
\newcommand{\seconds}{\unit{s}}
\newcommand{\micro}{\myfm{\mu}}
\newcommand{\micrometer}{\micro\mbox{\rm m}}
\newcommand{\Mdot}{\myfm{\dot M}}
%
%
%
\newcommand{\dgr}{\myfm{^\circ} }
\newcommand{\ddgr}{\mbox{\dgr\hskip-0.3em .}}
\newcommand{\mnt}{\mbox{\myfm{'}\hskip-0.3em .}}
\newcommand{\scnd}{\mbox{\myfm{''}\hskip-0.3em .}}
\newcommand{\hr}{\myfm{^{\rm h}}}
\newcommand{\dhr}{\mbox{\hr\hskip-0.3em .}}
%
%
%
%
%
%
%
\newcommand{\refindent}{\par\noindent\hangindent=0.5in\hangafter=1}
\newcommand{\figpar}{\par\noindent\hangindent=0.7in\hangafter=1}
%
%

\newcommand{\mybiblio}{\vspace{1cm}
		       \setcounter{subsection}{0}
		       \addtocounter{section}{1}
		       \def\junksec{References} 
 }

%
%
%

%
%
%
%
%

\newcommand{\vol}[2]{ {\bf#1}, #2}
\newcommand{\jour}[4]{#1. {\it #2\/}, {\bf#3}, #4}
\newcommand{\physrevd}[3]{\jour{#1}{Phys Rev D}{#2}{#3}}
\newcommand{\physrevlett}[3]{\jour{#1}{Phys Rev Lett}{#2}{#3}}
\newcommand{\aaa}[3]{\jour{#1}{A\&A}{#2}{#3}}
\newcommand{\aaarev}[3]{\jour{#1}{A\&A Review}{#2}{#3}}
\newcommand{\aaas}[3]{\jour{#1}{A\&A Supp.}{#2}{#3}}
\newcommand{\aj}[3]{\jour{#1}{AJ}{#2}{#3}}
\newcommand{\apj}[3]{\jour{#1}{ApJ}{#2}{#3}}
\newcommand{\apjl}[3]{\jour{#1}{ApJ Lett.}{#2}{#3}}
\newcommand{\apjs}[3]{\jour{#1}{ApJ Suppl.}{#2}{#3}}
\newcommand{\araa}[3]{\jour{#1}{ARAA}{#2}{#3}}
\newcommand{\mn}[3]{\jour{#1}{MNRAS}{#2}{#3}}
\newcommand{\mnras}{\mn}
\newcommand{\jgeo}[3]{\jour{#1}{Journal of Geophysical Research}{#2}{#3}}
\newcommand{\qjras}[3]{\jour{#1}{QJRAS}{#2}{#3}}
\newcommand{\nat}[3]{\jour{#1}{Nature}{#2}{#3}}
\newcommand{\pasa}[3]{\jour{#1}{PAS Australia}{#2}{#3}}
\newcommand{\pasj}[3]{\jour{#1}{PAS Japan}{#2}{#3}}
\newcommand{\pasp}[3]{\jour{#1}{PAS Pacific}{#2}{#3}}
\newcommand{\rmp}[3]{\jour{#1}{Rev. Mod. Phys.}{#2}{#3}}
\newcommand{\science}[3]{\jour{#1}{Science}{#2}{#3}}
\newcommand{\vistas}[3]{\jour{#1}{Vistas in Astronomy}{#2}{#3}}

\newcommand{\mybit}[1]{ {#1} }     
\newcommand{\leftb}{<\!\!} \newcommand{\rightb}{\!\!>}

%
   \title[Scaling up Tides in Numerical Models]
{Scaling up Tides in Numerical Models of Galaxy- and Halo-Formation} 
   \author[Boily, Athanassoula \&  Kroupa]{C.~M. Boily$^{1,\dagger}$, E. Athanassoula$^{2}$ \&
          P. Kroupa$^{3}$ \\
$^{1}$Astronomisches Rechen-Institut, M\"onchhofstrasse 12-14 Heidelberg, D-69120 Germany 
\\
$^{2}$Observatoire de Marseille, 2 Place LeVerrier, Marseille Cedex 47000, France 
\\
$^{3}$Institut f\"ur Theoretische Physik und Astrophysik der
Universit\"a Kiel, D-24098 Kiel, Germany \\ 
$^{\dagger}$Present address: Observatoire astronomique de Strasbourg, 11 rue de l'universit\'e, Strasbourg F-67000, France
}
   \date{Received September 15, 1999; accepted March 16, 2000}
   \maketitle

   \begin{abstract} 
 The purpose of this article is to show that when dynamically cold, dissipationless 
  self-gravitating systems collapse, their evolution is a strong
 function of the symmetry in the initial distribution. We explore with
 a set of pressure-less homogeneous fluids the time-evolution of
 ellipsoidal distributions and map the depth of potential achieved
 during relaxation as function of initial ellipsoid axis ratios. We
 then perform a series of $N$-body numerical simulations and contrast
 their evolution with the fluid solutions. We verify an analytic relation
 between collapse factor ${\cal C}$ and particle number $N$ in
 spherical symmetry, such that ${\cal C} \propto N^{1/3}$. 
We sought a similar relation for  axisymmetric
 configurations, and found an empirical scaling relation such that ${\cal C}
 \propto N^{1/6}$ in these cases. 
We then show that when mass distributions do not respect spherical- or 
axial-symmetry, the ensuing gravitational collapse deepens
 with increasing particle number $N$  but only
 slowly: 86\% of triaxial configurations may collapse by a factor 
of no more than  $40$ as $N\rightarrow\infty$. 
 For $N\approx 10^5$ and larger, violent relaxation develops  
fully under the Lin-Mestel-Shu instability such that numerical 
$N$-body solutions now resolve the different initial morphologies 
adequately.  
   \end{abstract}

%

\section{Introduction}  Cold, sub-virial distributions of  
      stars undergo  a phase of  gravitational focusing 
 during which binding  energy is redistributed between them 
 (Lynden-Bell's `violent relaxation' process, see Binney \& Tremaine
 1987 [$\equiv$ BT+87]).
 The equilibria established through this process show density 
profiles which, when 
averaged over spherical shells, approach a de Vaucouleurs 
law applicable to elliptical galaxies (van Albada 1982, McGlynn 1984). This 
has since motivated studies of galaxy and galactic halo formation 
 by some degree of gravitational relaxation (e.g. Henriksen \& Widrow 1997, Weinberg
 2001). A hands-on approach to this problem, free of geometric
 constraints, consists in integrating the equations of motion with 
   $N$-body numerical codes.  
A crucial step when applying results from $N$-body experiments to actual galaxies
and haloes consists in  bridging  the gap between simulation particle
numbers and the actual number of stars (or generally, mass elements) in
 galaxies, which still differ by  five orders of magnitude or
 more in present-day simulations of collisionless dynamics
 (Athanassoula 2000). It is therefore essential to establish the
 scaling of $N$-body results with particle number. 

The following  example in spherical symmetry 
brings the problem to focus. A  
   star at rest converges to the centre of gravity of a free-falling
 distribution of mass $M$ in an interval of time 
\begin{equation} \tff = \sqrt{ \frac{3\pi}{32 G <\!\!\rho(a,0)\!\!>
   } } \label{eq:tff} \end{equation}  
where $\leftb\rho(a,0)\rightb = 4\pi M/3 a^3 $ is the mean density inside the
star's initial radius, $a$. 
(The result holds when stars accrete at the origin, so that shell-crossing
 is suppressed, see Lynden-Bell 1973.) When the density profile is
flat initially, all 
stars converge to a point   in a free-fall time. 
The time-dependent gravitational potential along a radial orbit is   

\begin{equation}  \Phi(r,t) = - \frac{G M}{2 R }\left( 3 -
\frac{r^2}{R^2} \right)\ ; \   
\left.\nabla^2 \Phi\right|_R  = - \frac{G M}{R^3} 
\label{eq:potential} \end{equation} 
with $r<R$, the system radius, and double-differencing  \wrt $r$ at
fixed time yields a measure of the tidal field at $r$. 
We note that the tide is unbound as collapse proceeds and
$R\rightarrow 0$. In general one would not expect a flat density
profile on the scales of a galaxy but rather a heterogeneous or clumpy
matter distribution. Furthermore fragmentation modes develop on all
scales in homogeneous, cold distributions (Aarseth, Lin \& Papaloizou
1988, hereafter ALP+88). Bound clumps would survive violent relaxation 
 if their binding energy is high (van Albada 1982, Tsuchiya 1998). However 
 the above argument suggests that remnant structures 
would  be severely affected by the maximal potential depth experienced 
during collapse (here, when $R$ reaches a minimum). What constrains 
this maximum,  and how does it scale with particle number? The
survival of bound substructures, the mixing of orbits  and energy
exchanges between stars  all relate to constraints set on
(\ref{eq:potential}). 
 
We explore these questions both analytically, through an idealised
model of collapsing self-gravitating pressureless 
fluids, and with a set of numerical $N$-body 
calculations to bring out discreteness effects. 
 We find that while collapse simulations in spherical symmetry 
 reproduces theoretical expectations of vigorous infall (ALP+88), 
small departures from sphericity lead to much gentler
collapse.  Specifically, we perform $N$-body simulations of collapsing
spheroids covering four  decades in 
particle number $N$ (up to 16 millions) using an FFT code to integrate the equations of
motion. We show empirically  that during  the collapse of spheroidal 
distributions, the maximum gravitational  energy achieved 
scales with particle number as $\propto N^{1/6}$ and so becomes
infinite as $N\rightarrow \infty$.     
\mybit{Triaxial initial configurations offer no such scaling with particle 
number. We find from the pressure-free fluid calculations that the 
maximum gravitational 
 energy achieved depends sensitively on the initial axis ratios. We survey 
the parameter space of axis ratios on a mesh of 836 points integrated with 
high resolution to quantify energy maxima.} 
We then use this to constrain the increase in  potential energy of
self-gravitating finite-$N$ systems. Thus for instance, unless triaxial 
mass profiles are initially rounder than 
E1, their linear size \mybit{on average will
 contract by no more than a factor $O(20)$.} 

We cover the mathematics  in Section 2 before giving details of our 
simulations in the section that follows. The implications of
our study to scenarios of galaxy formation are
presented in the last section. 

\section{Analysis} Our overall objective is to find out the maximum 
potential depth haloes and galaxies  may attain during  
 violent relaxation. For constant initial density, equation 
 (\ref{eq:tff}) 
shows that the first caustic 
occurs at the unique free-fall time $\tff$. 
This would involve the whole of the galaxy, as opposed to
 a subset of stars. It  is therefore the profile of choice for our study. 
We justify this choice in part below and later with numerical modelling (see Section 4). 
 
\subsection{Results in Spherical Symmetry} We begin by considering in
more details the collapse of uniform density spheres. The results of
this subsection are those of ALP+88. 

The solution for radial infall of uniform spheres takes the  
parametric form

\begin{equation} \sin 2\eta + 2\eta = \left( \frac{8GM}{r(0)^3}\right)^{1/2}
  \times ( t - t_o ) \label{eq:eta} \end{equation} 
with $r(t) = r(0) \cos^2\eta, \ \eta(t=t_o) = 0$ and free-fall is
complete  when $\eta 
=\pi/2$. If we perturb the density profile so $\rho :
\rho_o + \delta\rho$, $\delta\rho > 0 $ (or, = 0) for $r < a$ (or, $ >a$),  
the free-fall time is now a function of position and no singularity forms. From
(\ref{eq:tff}) we write the new collapse time 

\begin{equation} \tff : \tff\times ( 1 -
 \delta\rho/\rho + O[(\delta\rho/\rho)^2] ) \equiv  \tff - \delta
\tff\, . \end{equation}   
Introducing \[ \eta(\tff) = \frac{\pi}{2} - \epsilon \]
 and inserting in  (\ref{eq:eta}), we find on truncating 
the  Taylor expansion to second order   in $\delta\tff$ 
\begin{equation} \epsilon^3 =   \left( \frac{8GM}{r(0)^3}\right)^{1/2}
\times \delta \tff \ . \label{eq:epsilon} 
\end{equation}  
Since the motion is  pressure-less, spherical mass shells  re-expand radially once
they have reached the centre. Those that originate outside $r = a$
meanwhile continue inwards. This spread in arrival time means that
the linear dimension of the sphere reaches a positive minimum. 
 This minimum is known in terms of the original system size and
$\delta \tff$ : 
   \begin{equation} {\cal C}^{-1} \equiv \frac{W(0)}{W(\tff)} = 
\frac{r(\tff)}{r(0)} = \cos^2\eta(\tff) =  \sin^2\epsilon
\approx \epsilon^2 \propto \left( \delta\tff\right)^{2/3}\ . 
\label{eq:factor} \end{equation} 
where ${\cal C}$ is the collapse factor, and $W\propto GM^2/r$ is the
system's gravitational energy. Note that the definition of ${\cal C}$ 
applies equally to non-spherical systems; the first equality in
(\ref{eq:factor}) is valid only for spheres. 

So far we have not specified the form of $\delta\rho$ in the region
$0 < r < a$  of the initial configuration, only that it be
positive. If furthermore $\delta\rho$ is a non-monotonic function of
position, shell-crossing will occur before any shell has reached the
centre. This would contribute to smooth out fluctuations by orbit
mixing, but would not affect infall of the system as a whole. In the
case of a point-mass realisation of a uniform density stellar system, 
  discreteness introduces  
 Poissonian noise so that  $\delta\rho/\rho \propto 1/\sqrt{N}$,  with $N$ 
the total particle number. The ratio (\ref{eq:factor}) now scales with  
the inverse one-third power of $N$, or 

\begin{equation} 
{\cal C} \propto N^{1/3}\ . \label{eq:scaling} \end{equation}  
 The density perturbations $\delta\rho > 0$ leading to (\ref{eq:scaling}) are 
Jeans-type fragmentation modes of instability: the enhanced gravity pulls in 
the matter which condenses faster at the origin, as in the classic Jeans 
condensation of star-formation studies. 
 The scaling relation (\ref{eq:scaling}) recovers the solution for a cold Newtonian fluid in the limit $N\ra\infty$, for which $r(\tff)\ra 0$. Systems with small particle numbers 
experience relatively larger density fluctuations, which act as seeds for 
fragmentation modes contributing to halt radial infall. Infall stops once orbit crossing occurs in the centre, as encapsulated by the spread in radial 
collapse time (\ref{eq:factor}). 

The above analysis would not apply to initially cuspy 
profiles, since in that case shell-crossing takes place immediately at the
centre. Analysis with shell-crossing 
is beyond our scope. However if we view a peaked profile as a perturbed,
uniform-density distribution, where a large-amplitude perturbation is
necessary to 
distinguish it from Poissonian noise, the spread in free-fall times,
$\delta\tff$,  would then be even larger, so presumably (\ref{eq:factor})
is minimised for initially flat profiles. 

\subsection{Non-Spherical Collapse} 
\subsubsection{Small deviations} Pressure-less, self-gravitating 
 oblate or prolate
structures collapse first down the shortest axis (Lynden-Bell 1964, Lin, Mestel \& Shu
1965). Hence perturbations breaking the spherical symmetry while
preserving internal density homogeneity grow in time. Writing 

\begin{equation} \boldv{r}(t) = r(0) \cos^2\eta \,\, \boldv{\hat{r}} + \boldv{\xi} \label{eq:surface} \end{equation} 
where $\eta(t)$ is known from (\ref{eq:eta}) and 
$\boldv{\xi}(r,\theta,\phi,t) = $ vector 
displacement of the spherical surface
 \wrt the radial direction $\boldv{\hat{r}}$. 
Lagrangian linear analysis
shows growth rates for these modes (cf. eq. [37] of ALP+88)

\begin{equation} \frac{\rmd^2 \boldv{\xi}}{\rmd t^2} = \frac{4\pi}{3} 
\frac{G \rho_o}{(\cos^2\eta)^{-3}}\frac{n-1}{2n+1}\boldv{\xi} \label{eq:xi} \end{equation} 
where $n > 1$ is
the principal number of an angular decomposition of the displacement
in spherical harmonics ($n = 1$ corresponds to a homogeneous radial
contraction). Thus the \rhs in the above equation is positive and $|\boldv{\xi}|$ becomes larger  in time. 
Since (\ref{eq:xi}) is linear in $\boldv{\xi}$, angular and radial components of each mode (or, value of $n$) grow in time at the same rate. We may solve numerically for 
$|\boldv{\xi}|$ as a function of time using (\ref{eq:eta}). However an approximate solution is found immediately if we note that for collapse in spherical 
symmetry the time-averaged square cosine, 

\[ \leftb \cos^2\eta \rightb = \half\,\]
gives a mean ratio $r(t)/r(0) = 1/2$ averaged over $\tff$.  
Substituting this in (\ref{eq:xi}) and writing $|\boldv{\xi}| = \xi(t)$ we
find on integrating 
\begin{equation} \xi(t) = \xi_o \, \exp\left( \sqrt{8\,\frac{4\pi}{3}G\rho_o\frac{n-1}{2n+1} } [t-t_o] \,\right) \, \label{eq:xi.of.t} \end{equation}  
Thus high-order $n\gg 1$ modes grow faster with increasing $t-t_o$.
 
At the time when $\xi(t) \simeq r(t)$, the linear size of the  
system may be compared with (\ref{eq:factor}) in order to 
determine which type of perturbations develop 
the fastest for a 
given particle number. For this purpose  we truncate to third 
order a Taylor expansion \wrt $t-t_o$ of (\ref{eq:xi.of.t}) 
in the limit $\eta\ra \pi/2\, ({\rm i.e.}, t-t_o\ra\tff)$. 
In the case of Poissonian fluctuations, we find from (\ref{eq:factor}) and 
(\ref{eq:scaling}) that the fragmentation modes in spherical symmetry 
 develop faster than surface modes (\ref{eq:surface}) when 

\begin{equation} N^{1/3} \ltabout  \left( \frac{\pi}{2} \sqrt{\frac{n-1}{2n+1}} + \frac{\pi^2}{2}\frac{n-1}{2n+1} + 1 \right)^{-1} \left(\frac{\xi_o}{r[0]} \right)^{-1} \label{eq:n} \end{equation} 
For a particle realisation of a uniform-density sphere, the statistics 
will be Poissonian. The surface mode should initially rise above the
noise level to be effective. We therefore set 

\[ \frac{\xi_o}{r(0)} = \frac{1}{\sqrt{N}}\, , \] 
 from which we find a critical particle number,  

\begin{equation} N_c^{1/6} = \frac{\pi}{2} \sqrt{\frac{n-1}{2n+1}} + \frac{\pi^2}{2}\frac{n-1}{2n+1} + 1\, ,   \label{eq:nc} \end{equation} 
such that for $N\gtabout N_c $ fragmentation modes (`clumpiness') 
halt radial infall 
 before $\sqrt{N}$-seed surface modes have developed fully and led to 
pancaking. 
 If we make $n \gg 1$ we compute the maximum value possible for  $N_c$: 

\begin{equation} N_c \approx 4.6^6 \simeq 9475 \ . \label{eq:nc2} \end{equation}
Otherwise said, this $N_c$ is the largest possible particle number
 for which discreteness effects (noise) may significantly distort the
 flow of a collapsing sphere through a pancaking mode. 

 The results (\ref{eq:nc}) and (\ref{eq:nc2}) apply to initially 
small amplitude deviations from spherical symmetry. 
For sufficiently large deviations from sphericity, 
linear analysis shows that small particle number simulations may yet reproduce 
the pancaking collapse of cold fluids (Lynden-Bell 1964, Lin, Mestel \& Shu 1965). Consider for example an axisymmetric displacement of 
amplitude  $\xi_o \simeq r(0)/4$ mapping a sphere to a spheroid of 
 aspect ratio 3:4. For this case 
we compute from (\ref{eq:n}) $N\ltabout 4^3 / (4.6)^3 \sim 1$.
Thus for axisymmetric cold distributions, of initial aspect ratio $ < 3/4$, 
any sensible simulation particle number will adequately reproduce 
 the Lin-Mestel-Shu flow. 
The time-evolution of axis-ratio of  collapsing triaxial systems with $N = 10^5$ particles  performed by 
 Hozumi et al. (1996) shows growth of surface modes (pancaking) 
in agreement with (\ref{eq:n}). The initial 
axis ratios of their systems were $\xi_o / r(0) \simeq 0.01 $ and $ 0.005$, or three times the Poisson noise level for this number of particles. 

 We stress that {\em uncorrelated} Poisson noise of a uniform spherical distribution
 is not sufficient in itself to lead to appreciable 
flattening during infall, for any particle number. Thus  the scaling 
(\ref{eq:scaling}) is well recovered from simulations with as few as 
$N \sim 10^2$ particles 
 (see ALP+88, Boily et al. 1999), therefore (\ref{eq:nc})
 does not invalidate the interpretation of previous  
studies of small-$N$ collapse simulations in terms of one-dimensional radial motion 
(e.g. van Albada 1982, Aguilar \& Merritt 1990, Cannizzo \& Hollister
 1992).

\begin{table*}
\centering 
\caption{Axes and velocity components for a triaxial pressureless configuration with initial axes $(a_1,a_2,a_3) = (1,\half,\frac{1}{3})$ for different 
values of the parameter $\epsilon$. The time $t_{max}$ refers to the moment 
when the configuration is most compact (maximum $W$). Integration ended at $t_{\rm final} = 4.398$ model units.}
\begin{tabular}{crccccccc}
$\epsilon$ &  $t_{max}$ & $W_{max}$ & $a_1$ & $\dot{a}_1$ & $\tau$ & $\tau_v$ & $a_1$ & $\dot{a}_1$ \\
           &            &           & $\div 10^{-3}$ &            &          & &  (at $t_{\rm final}$)&  (at $t_{\rm final}$) \\
\hline  \\
$1\times10^{-3}$ & 1.934 & $6.21$ &  1.780 & 2.291 & 0.257 & 0.125 & 0.358 & -0.008 \\
$5\times10^{-4}$ & 1.933 & $6.31$ &  0.954 & 2.291 & 0.259 & 0.169 & 0.372 & -0.044 \\
$1\times10^{-4}$ & 1.930 & $6.40$ &  0.169 & 2.291 & 0.262 & 0.206 & 0.377 & -0.074 \\
$5\times 10^{-5}$ & 1.931 & $6.41$ & 0.014 & 2.291 & 0.262 & 0.212 & 0.381 & -0.075 \\
$1\times10^{-5}$ & 1.930 & $6.42$ &  0.025 & 2.291 & 0.263 & 0.213 & 0.382 & -0.076 \\
$5\times 10^{-6}$ & 1.930 & $6.42$ & 0.024 & 2.291 & 0.263 & 0.213 & 0.383 & -0.075 \\ 
$1\times10^{-6}$ & 1.930 & $6.42$ &  0.025 & 2.291 & 0.263 & 0.214 & 0.381 & -0.079 \\ 
\end{tabular}
\label{tab:testorbit}
\end{table*}

\subsubsection{Large deviations: ellipsoidal figures}
An uniform-density ellipsoid collapses down the minor axis  before
major-axis collapse is complete, followed by re-expansion when the
fluid is also pressure-less. Since all orbits are synchronous in a
homogeneous distribution, phase-mixing is minimal. Minor-axis cyclic
motion continues while the ellipsoid collapses down the major axis,
until it too rebounds while the minor-axis assumes a finite value. 
For a perfect fluid, such cycles of collapse/expansion may repeat
themselves without loss of cohesion. For a fluid made up of stars
however, the stars exchange kinetic energy and phase along their
orbits, causing damping.  In the case where the 
 system is initially oblate spheroidal, axes $a_1 = a_2 > a_3$, 
Boily et al. (1999) found in $N$-body simulations the 
time-evolution of the aspect ratio after plane crossing to be approximately 
given by 

\begin{equation} \frac{a_3(t)}{a_1(t)} = \frac{a_{3}(0)}{a_{1}(0)}\left(\frac{a_1(t)}{a_{1}(0)}\right)^{-1/3} \label{eq:adiabat} \end{equation} 
i.e. the spheroid becomes rounder as collapse continues and the major
axis $a_1(t) \ra 0$. In practice only a few minor-axis oscillations are
detected before phase mixing and violent relaxation lead to
equilibrium with little or no streaming pattern. We want to establish
for triaxial initial configurations a constraint on the collapse
factor ${\cal C}$ in (\ref{eq:factor}) by solving the equations of
motion for a pressure-less ellipsoidal fluid integrated over a
timescale for complete relaxation suggested by $N$-body simulations. The 
motion of a triaxial uniform ellipsoid, of axes $a_1>a_2>a_3$, is
governed by a set of harmonic equations (BT+87, Table 2.1) 

\begin{equation} \frac{\rmd \tilde{v}_i }{\rmd \tilde{t}} = -\nabla_i 
\Phi(\boldv{\tilde{x}}) = - \frac{\tilde{x}_i}{\tilde{x}^3_1} A_i
(\boldv{a}) \label{eq:ellip1} 
\end{equation} 

\begin{equation} \frac{\rmd \tilde{x}_i }{\rmd \tilde{t}} =
\tilde{v}_i \label{eq:ellip2} 
\end{equation} 
where $\tilde{x}_i(\tilde{t}) = a_i(\tilde{t})/a_{i}(0)$,
$\tilde{v}(\tilde{t}) = \rmd\tilde{x}_i/\rmd\tilde{t}$ are
dimensionless functions of the dimensionless time $\tilde{t} \equiv
t/\tff$. The coefficients $A_i(\boldv{a})$ are known from potential
theory. For instance we have 

\[ A_1 (\boldv{a}) \equiv 2\frac{a_2(0) a_3(0)}{a_1^2(0)}\,
\frac{F(\theta,k) - E(\theta,k)}{k^2\sin^3\theta} \] 
with similar definitions for $A_2, A_3$, and 

\begin{equation} k(\tilde{t}) \equiv \left(
\frac{a_1^2-a_2^2}{a_1^2-a_3^2}\right)^{1/2}\ ;\ \theta(\tilde{t}) \equiv \cos^{-1}
( \frac{a_3}{a_1}) \ . \label{eq:defktheta} \end{equation} 
Here $F,E$ are incomplete elliptical integrals (see BT+87 for
details). We may identify the most relevant configurations by
inspecting the gravitational energy. The self-gravitating potential
energy $W$ is known for uniform ellipsoids from 

\begin{equation} W(\boldv{a},t) = - \frac{3}{5}\, \frac{M^2}{a_1(t)}\,
\frac{F(\theta,k)}{\sin\theta} \label{eq:W} \end{equation} 
with the definitions (\ref{eq:defktheta}). The energy $W$ in
(\ref{eq:W}) diverges when the ellipsoid collapses to a rod (or
spindle) which is the case when $\boldv{a} \rightarrow (a_1,0,0)$. It
remains finite when two axes are non-zero, which includes collapse to
a disc. Presumably the force and tidal fields are maximised when
ellipsoids develop spindles, together with $W$. To constrain the tidal 
field, it is therefore sufficient to determine when an ellipsoid forms 
a spindle, or, generally, what maximum value $W$ may reach during
evolution. We were not able to determine these analytically and have
resorted to a numerical integration of the equations of motion. We
found it useful to introduce the parameter $\tau$ defined as 

\begin{equation} \tau \equiv \frac{a_2-a_3}{a_1-a_3} \geq 0 \
. \label{eq:deftau} \end{equation}
Thus axisymmetric prolate spheroids ($a_2=a_3$) all have $\tau = 0$,
whereas axisymmetric oblate spheroids ($a_1=a_2$) have $\tau =
1$. Triaxial structures assume intermediate values.

%
%
   \begin{figure*}
\setlength{\unitlength}{2cm} 
\begin{picture}(6,9)(0,0) 
	\put(.3,1.5){ \epsfysize=5.0in
		    \epsffile[100 200 450 600]{null.ps} 
                  }
\end{picture} 
      \caption{Maximum potential energy $W_{max}$ achieved by
         pressure-less homogeneous ellipsoids of axes $a_1 = 1 > a_2 >
         a_3$ collapsing from rest. $W$ has been normalised to its
         initial value. The $W_{max}$ surface is projected in the
         $a_2-a_3$ plane and three contours are shown, $W/W_o = 12, 18$
         and 24. The maximum values displayed have been truncated to 25.}
         \label{fig:Wmax}
   \end{figure*}

\section{Solutions for Pressure-less Ellipsoids.}
\label{sec:pressurelessellipsoids} 
\subsection{Method \& Tests} 
For given axes $\boldv{a} = (a_1,a_2,a_3)$ we may integrate
(\ref{eq:ellip1}) and (\ref{eq:ellip2}) subject to the initial
conditions $\tilde{x}_i(0) = 1, \tilde{v}_i = 0.$ Note that the
equations are singular when any of the axes vanishes. To integrate
through such singularities we enforce time-symmetry by reversing the
flow: $\tilde{v}_i \rightarrow - \tilde{v}_i$ whenever $\tilde{x}_i
\leq \epsilon$, with $\epsilon$ a free parameter. We set up a 
fourth-order Runge-Kutta integrator (Press et al. 1992) and varied
$\epsilon$ 
from $ 10^{-3}$ down to $1\times 10^{-6}$ without appreciable 
differences in the integrated global quantities \mybit{such as maximum $W$ and 
time (see Table~\ref{tab:testorbit}). However details of the fluid configurations converged to good accuracy only when $\epsilon \approx 5\times 10^{-5}$ or 
less. To quantify the quality of orbit-integration, we computed both axial 
lengths and velocity components for the representative case where 
$(a_1,a_2,a_3) = (1,\half,\frac{1}{3})$;  
 then initially  $\tau = \frac{1}{4}$  from (\ref{eq:deftau}). We computed 
a similar quantity $\tau_v$ from the velocity components $(\dot{a}_1,\dot{a}_2,\dot{a}_3)$ which we evaluated  at  $t = t_{max}$, when $W/W(0)$ is maximum.  

The results are listed in Table~\ref{tab:testorbit} for various values of $\epsilon$. By comparing the runs of individual  components $(a_1, \dot{a}_1)$ and those of $(\tau, \tau_v)$ 
 with decreasing $\epsilon$, we may conclude that both 
 geometric and velocity ellipsoids vary little with $\epsilon$.  
  This does not hold for individual components, such as major axis, 
$a_1$. This last quantity must reach a minimum a few times 
  larger than $\epsilon$ to ensure that the dynamics is resolved properly. 
 Drawing from the results in the Table, 
 we set $\epsilon = 2\times 10^{-5}$ or $1.5\times 10^{-5}$ 
in all our integrations as a minimal condition to accurate integration.} 

\mybit{In conjunction with the value of $\epsilon$, the choice of timestep is 
crucial: we adopted a time-adaptative scheme tailored to the instantaneous free-fall 
time (\ref{eq:tff}) at each step. This allowed us to resolve in time increases in potential 
energy by factors up to $\approx 340$, while keeping errors below the 1\% level (though not for axisymmetric systems, see below). 
}

When integrating equations (\ref{eq:ellip1}) and (\ref{eq:ellip2}), 
care must be
 taken that the indices $(1,2,3)$ are circulated between each axis to
identify the current major and minor axes properly, and allow the
correct evaluation of the force field. All integrations were done in 
three dimensions but we found it necessary to enforce symmetry in the
potential  when treating  initially axisymmetric configurations in order to 
prevent large numerical errors. With enforced symmetry,
 we computed  correctly the growth of singular axisymmetric potentials for a 
collapse factor reaching $\approx 40$. \mybit{As a test, we integrated through 
 the singularities formed through major-axis collapse of axisymmetric spheroids with 
initial axes $(1,\half,\half)$ and $(1,1,\half)$. The error in 
binding energy at the end of 
 integration reached $5\%$ and $6\%$ in each case respectively, 
which we take as reference later when sampling the space of triaxial 
 initial configurations. 
Note that these errors were accrued during a single integration through singularity: the 
slow divergence of $W\rightarrow \infty$ as axes vanish makes agreement with theory 
 intractable, however the important quantity for these cases must be the time 
$t$ when $W$ diverges (since the spindle morphology is known), 
which could be evaluated to high accuracy.  }
  
 In order to determine the sensitivity of our integration scheme to the symmetry of the system, we 
  integrated a triaxial configuration with axes $= (1,0.99, \half)$, i.e. nearly oblate 
axisymmetric. 
In this case integration yielded a maximum collapse factor of $\approx 26$ and 
 total error accrued for the system binding energy of 
$\approx 0.5\%$ with the same set-up used for the strictly axisymmetric case discussed 
before. We concluded from this that the integrator resolves relatively small 
 departures from axisymmetry, of the order of one percent in axial ratios; and that these 
small departures are sufficient to avoid spindles and large energy errors, while still 
collapsing by appreciable factors. 

We have monitored the three axes $\tilde{x}_i$ as a function of time to
determine whether a spindle forms, which, in view of our
approximations, is the case when 

\begin{equation} \boldv{a} \rightarrow (a_1,\epsilon,\epsilon)\ ; \
a_1 > \epsilon\ :\ \ {\rm definition \ of \ a \ spindle.}
\label{eq:defspindle} \end{equation} 
If this occurs, then a self-gravitating body may assume \mybit{an even  larger} 
collapse factor ${\cal C}$ defined in (\ref{eq:factor}) since \mybit{our scheme does not 
resolve the dynamics below that scale. The identification of configurations of high 
potential energy is made difficult due to the slow divergence of $W$ with vanishing 
minor axes $a_2,a_3$ (spindle), or all three axes simultaneously (spherical case). 
By contrast, the case when only one axis vanishes (by definition not the major axis) is integrable to high 
accuracy.  We therefore 
computed the logarithmic averaged length $l \equiv (a_1\, a_2\, a_3 )^{1/3}$,
 which we used  together with (\ref{eq:defspindle}) to ensure that 
the maximal potential energy computed occurred at the same time as 
  the minimum of $l$. As a final precaution, we 
rejected any integrated solution that accumulated errors in binding energy exceeding 10\%, 
in view of our tests with axisymmetric configurations.}  
When (\ref{eq:defspindle}) does not occur, the collapse
factor  reaches a finite maximum: this maximum is then the limit
any $N$-body collapse calculation for this initial geometry may reach, since discreteness effects will 
only increase the growth of kinetic energy and slow down collapse.

   \begin{figure}
\setlength{\unitlength}{1cm} 
\begin{picture}(6,7.)(0,0) 
	\put(1.25,2.25){ \epsfysize=6.0cm
		    \epsffile[100 200 450 600]{null.ps} 
                  }
\end{picture} 
      \caption{Integrated distribution (in \%) for ellipsoidal 
          pressure-less uniform fluids as function of the maximum 
gravitational energy achieved during infall.  For each 10\% interval, we give the
         mean shape parameter $\leftb\tau\rightb$ (eq. [\ref{eq:deftau}])  and its standard deviation 
 evaluated from the initial conditions.} 
         \label{fig:distribution}
   \end{figure}

   \begin{figure}
\setlength{\unitlength}{1cm} 
\begin{picture}(6,15.)(0,0) 
	\put(1.0,10.25){ \epsfysize=6.0cm
		    \epsffile[100 200 450 600]{null.ps} 
                  }
	\put(1.0,2.5){ \epsfysize=6.0cm
		    \epsffile[100 200 450 600]{null.ps} 
                  }
\end{picture} 
      \caption{Top panel: scatter plot showing the maximum $W$ achieved as function of the initial 
shape parameter $\tau_0$. Lower panel: comparison between the initial morphology  (measured by $\tau$, cf. eq.~\ref{eq:deftau}) and the 
morphology when $W$ is maximal. The integrated distribution of $\tau$ is shown in each case.  This demonstrates a drift towards lower $\tau$ during infall.}
         \label{fig:taudist}
   \end{figure}

\subsection{Results} 
Anticipating the results from $N$-body calculations of 
Section~\ref{sec:spheroids}, we integrated (\ref{eq:ellip1}) and (\ref{eq:ellip2}) up to 
$t = \tff\, (\tilde{t} = 1)$, corresponding to one free-fall time
(\ref{eq:tff}) in spherical symmetry. We then explored the parameter
space of $(a_{2}(0),a_{3}(0))$ by sampling the parameter $\tau$ uniformly in the 
interval [0,1], for a total of 900 pairs $(a_2,a_3)$, 
while fixing the major axis to an initial value $a_{1}(0) = 1$. We
evaluated (\ref{eq:W}) and kept the largest value, $W_{max}$, found during
integration. The results are displayed on Fig.~\ref{fig:Wmax}. The
bottom panel graphs the surface of maximum energy for all pairs
($a_{2}(0), a_{3}(0)$). It is striking that large islands exist where 
Max($W) \sim O(10) $, whereas all axisymmetric configurations with 
$\tau = 1$ or 0 (i.e., $a_2 = 1$ and $a_2= a_3$, respectively) must develop 
spindles and infinite $W$. The vertical axis has been capped to
$W_{max}/W(0) = 25$ for clarity. Larger values were not prohibited in 
the course of integrating numerically. The presence of fragmented regions 
 with large $W$ in the plane $(a_2,a_3)$ are indicative of the
formation of spindles or very compact configurations, in the sense of
our equation (\ref{eq:defspindle}). 
We note the presence of islands of as few as a single point were
$W_{max}/W(0) \geq 25$, suggesting a complex topology.  Details of the topology of the 
energy surface $W$ are not important to the main argument and will not 
be pursued further. 

Cases where spindles formed from initially triaxial configurations
turn out to be exceptional. For these cases, a repeat of the
integration with a smaller $\epsilon$ lead to larger Max($W$) at the
time  the spindle formed, requiring careful step-wise integration 
over a small time interval. The remainder of the integrated solution, 
however, was left largely unchanged, giving confidence that
singularities were correctly identified and cured. \mybit{As stated earlier, 
we rejected all runs that accumulated errors in binding energy through full time-integration 
 exceeding 10\%: in total 
some 62 (i.e. 7.1\%) of all initial configurations were rejected on this basis. 
 A graph of $\delta E/E_o $ vs Max($W/W(0)$) showed 
 no clear trend with increasing Max($W/W(0)$) for these 62 cases, 
 as would have been the case  if a single integration 
through singularity at high density was accountable for  the bulk of the error budget: instead, 
 large numerical errors develop owing to repeated integration through singularities, which 
may occur in rapid succession if the initial configuration is significantly non-axisymmetric.  Indeed of the 62 cases 
with significant energy errors, 37 initially had axis ratios $a_3/a_1 = 1/10 $ or lower; two more  showed 
near sphericity, with $a_2/a_1$ and  $a_3/a_1 > 0.999$. The remaining 23, 
however, showed no peculiarities  in their 
initial values, or in the maximum $W$ computed. They were, nevertheless, 
 left out of the analysis.} 

From Fig.~\ref{fig:Wmax} we may quantify the fractional area in
($a_2,a_3$) leading to large potential energy and collapse factor. The
inset gives the projected isocontours in the parameter space. If we
assume fair sampling, then the total area covered by the highest-level 
contours is estimated easily using a rectangular grid to cover the 
contoured area. In this way we compute a net fraction of $\approx
30\%$ of the total area exceeding an increase in $W/W(0)$ of 25. Note
that practically all triaxial configurations with $a_2 \ltabout 0.4$ and $a_3 \ltabout 0.2$ 
likely develop large $W$. These considerations are quantified more
accurately by sorting $W_{max}/W(0)$ in increasing order for all pairs
$(a_2,a_3)$. This gives an integrated distribution of  
 collapse factor ${\cal C}$. 
Figure~\ref{fig:distribution} plots the \mybit{integrated fraction}
 of 836 solutions 
as function of $W_{max}/W(0)$. Two-thirds of these solutions reached a 
collapse factor ${\cal C} < 22.4$, and 86\% have ${\cal C} < 40$. 
We sought a correlation between the morphological parameter $\tau$ and the 
maximum potential reached. On Fig.~\ref{fig:distribution} we also plot the 
mean $\tau$ computed at $t = 0$ 
for each ten-percentile interval, in increasing order of $W_{max}/W(0)$. 
We find a broad trend such that ellipsoidal initial conditions with 
larger $\tau$ tend to collapse to deeper potentials. We may identify
for the first of these bins, which has $<\tau> = 0.24$, the broad
low-$W_{max}$ valley seen on Fig.~\ref{fig:Wmax}.
\mybit{The non-monotonic relation of $<\!\!\tau\!\!>$ with $W_{max}/W(0)$ 
may be guessed at if we look at 
a scatter plot of this quantity versus $\tau$ directly, as shown on the top panel of Fig.~\ref{fig:taudist}. 
 The points clustered to the bottom left corner of the graph clearly reflect the trend of small $\tau$ 
to yield small $W_{max}/W(0)$. The broad trend we compute for $<\!\!\tau\!\!>$ 
can be guessed from shifting a 
horizontal ruler vertically up the $W$ axis: there is a suggestion of a gap in the data which 
 account for the dip in $<\tau>$ in the 50-60\% interval 
(when $W_{max}/W(0) \approx 20$).  
Our conclusions concerning the significance
 of this  gap must be 
moderated by the large  deviations about the mean values.  A more robust
 signature of dynamical evolution is a shift of the distribution of 
$\tau$ towards lower values  during infall. 
 This may be measured by computing the shape parameter $\tau$ at the time when $W$ is 
maximum, and compared  
to the initial value, $\tau_0$, for a given configuration. The result is shown on the bottom panel of 
Fig.~\ref{fig:taudist}. At maximum $W/W(0)$, the mean $\tau = 0.29$ approximately, down from 0.50 for the initially  
uniform distribution. By the definition (\ref{eq:deftau}), this implies a roundening of the two minor axes, 
in the same fashion as occurs for axisymmetric spheroids, that is $a_3/a_2$ increases during 
infall  (see Boily et al. 1999 for a discussion of this issue). 
  If the systems were allowed to virialise, as will happen in an $N$-body calculation allowing 
orbit mixing and violent relaxation, in contrast to the fluid model presented here, 
the effect measured 
during infall of the increasing ratio $a_3/a_2$, so enhancing axisymmetry, 
would be offset by the onset of radial orbit instability, 
which is known to enhance triaxiality (see e.g. Aguilar \& Merritt 1990). We have not addressed here 
the question of which effect would prevail in determining the equilibrium of haloes or galaxies. This issue 
would require more realistic density profiles than used here and is beyond the scope of the present paper.}

\subsection{Summary} 
We sum up the results for pressure-less self-gravitating ellipsoidal
collapse as follows: 1) The bulk of initially triaxial figures does
not reach a collapse factor ${\cal C} = W/W(0)$ 
exceeding 40. We find that 86\% of the 836  
configurations examined 
collapsed to smaller values; 2) Triaxial figures that
collapse by small  factors (small increase of $W/W(0)$) tend to 
have preferentially small values of the parameter $\tau$ (i.e., are 
more axisymmetric, see Fig.~\ref{fig:distribution}); 
 3) the axis ratio $a_3$ to $a_2$ increases during infall, leading to more 
axisymmetric configuration at maximum collapse (Fig.~\ref{fig:taudist}); 
 4) since no orbit mixing or fragmentation 
mode develop in the fluid solutions, which would contribute to \mybit{boost velocity dispersion and stop infall}, 
 we deduce that the collapse factors ${\cal C}$ obtained are absolute maxima. 

In order to apply correctly these results to galaxies and haloes, we
must first quantify the impact of discreteness effects of finite-$N$
systems as discussed in Section 2. We proceed empirically with $N$-body 
numerical calculations. We turn first to the task of reproducing the 
theoretical expectation (\ref{eq:scaling}) for spherically symmetric
systems. This is followed by a set of calculations of non-spherical
initial configurations, from which we seek trends with particle number 
to compare with the results for perfect fluids of
Fig.~\ref{fig:Wmax}. 
 
\section{$N$-body calculations: Set-Up and Tests} 
\subsection{Numerical code and units} \noindent
 Our intention is to cover as wide a range in particle number as possible 
to seek out correlations applicable to galaxies and haloes. The nested-grid
code {\sc superbox} was used (Fellhauer et al. 2000). 
This is an FFT Poisson solver with Cartesian grids, of  uniform 
resolution for each cell of a given grid. There are three grid levels,
each within one another, thus enabling enhanced resolution where it is
required. This is a crucial feature for the problem at hand, since the 
structures collapse by large factors and the density increases
accordingly. For computational 
purposes the units were chosen such that the total mass and initial radius 
of the system = 1, however the gravitational constant $ G = 2$. The free-fall 
time (\ref{eq:tff}) for uniform spheres is therefore 

\begin{equation} \tff = \frac{\pi}{4} \simeq 0.785398 ... \,
. \label{eq:newtff} \end{equation} 
A version of the code 
was set up where the integration timestep of a leap-frog scheme 
 scales down with the inverse square root of the  
local density maximum during the simulation, in order that the timestep 
$\delta t$ remains in the same ratio to the instantaneous dynamical
time (or, $\tff$ evaluated from [\ref{eq:tff}] but with
$\leftb\rho\rightb$ now 
the density at time $t$). The  three levels 
of resolution allowed by the code were set such that when the system reaches 
a minimum size, the number of particles per cell is of the order of a
few, 
hence interactions between particles are resolved approximately 
on the smallest scales. We found in practice that grid resolution smoothes 
out forces between particles and hence leads to a less deep potential
minimum at the centre of gravity. 
 As we show below, the net effect of gridding can be easily 
 brought to low error levels. 

   \begin{figure*}
\setlength{\unitlength}{1in} 
\begin{picture}(4,3)(0,0) 
	\put(-.40,-1.33){ \epsfysize=5.50in
		    \epsffile[100 200 450 600]{null.ps} 
                  }
\end{picture} 
      \caption{Orbits of four Lagrangian mass shells during the infall of a
      uniform-density sphere. The shells enclose 10, 30, 50 and 80 \%
      of the total mass. Results for 1 million 
      particle simulations with low (left) and high (right) resolution 
	grids are displayed. The dashed line is the Kepler
      scaling 
	of orbits, $ r^3 \propto (\tff-t)^2$. At constant internal mass, each 
	orbit converges asymptotically to Kepler scaling under adequate
	resolution.}
         \label{fig:kepler}
   \end{figure*}
\subsection{Tests in spherical symmetry}
We carried out checks of our set-up in spherical symmetry before conducting a 
survey of particle number and geometry. 
 All simulations of violent relaxation proceed from zero velocity initially. 
Our results 
are summarised in Table~\ref{tab:Spheres}. The analysis of Section 2.1 
suggests a clear relation (Kepler's) between system radius and the time 
interval 
$\delta\tff$ before collapse. From (\ref{eq:epsilon}) and (\ref{eq:factor}), 

\[ r(\tff-t ) \approx r(0) \sin^2 \epsilon \propto (\tff-t)^{2/3}\ .\]
We recover this relation for large particle number simulations shown on 
Fig.~\ref{fig:kepler}. This graphs the evolution of four constant mass
(Lagrangian) shells for a uniform density sphere of one 
 million particles. The initial radius $r(0) = 1$. 
 Because the free-fall time is independent of position, the Lagrangian  
radii must remain in the same relative ratio to one-another; each must converge
to the Keplerian regime near full collapse. The time of collapse $(t-\tff = 0)$
 is off-scale on the right-hand 
side of the logarithmic abscissae  on the figure. 
Two set-ups are illustrated, of low- 
(left-hand) and high-resolution (right-hand) grids. 
 The linear high-resolution achieved, $l = $ grid size / number of
cells $= 0.05 / 64 = 0.0008 $, is still 
large when compared with 
the mean inter-particle distance $l_{int}$ expected from (\ref{eq:scaling}),
  \[ {\rm \small Volume\ at\ bounce =}\, N\, (l_{int}/2)^3 = \hfill \]
\[\hfill {\rm \small Initial\ volume\ \div N = }\, r(0)^3/N, \] 
 or 
$ l_{int}  \approx 2\times N^{-2/3} = 2\times 10^{-4}$. If we  count the average number of particles in a (cubic) cell at the bounce, we find 
$\sim (l/4 l_{int})^3 \sim O(1)$ particles. 
  The mass distribution is therefore well sampled, and as a result
both the constant 
ratios of Lagrangian radii and their match of the Keplerian relation are 
reasonably well recovered. By comparison, for runs with 
reduced  number of mesh points, from 128 to 32, we find $\approx 100$ 
 particles in each cell at the bounce. 
Poor resolution of the mass distribution leads 
to artificial deviations from the Keplerian tracks (see left-hand panels 
on Fig.~\ref{fig:kepler}). The Lagrangian radii  spread out,  
 which results in shell crossing at the centre while the outer shells 
continue to fall in: this causes artificial orbit mixing and a gentler
collapse. 

   \begin{figure*}
\setlength{\unitlength}{1cm} 

\begin{picture}(6,7)(0,0) 
	\put(.0,0.25){ \epsfysize=7.0cm
		    \epsffile[100 200 450 600]{null.ps} 
                  }
\end{picture} 
      \caption{(A) Collapse factor ${\cal C}$ 
versus particle number $N$. Uniform spheres follow the scaling
      relation $\propto N^{1/3}$ (solid line) well when the grid mesh resolves 
       particles individually (black circles). The open circles 
are results obtained by ALP+88 with a direct-summation code. Crosses
were taken from Theis \& Spurzem (1999) for Plummer models using the 
direct-integration code NBODY6++. Their data follows the same
power-law as uniform distributions (dotted line). }
         \label{fig:scaling}

\setcounter{figure}{4} 

\setlength{\unitlength}{1cm} 

\begin{picture}(6,7)(0,0) 
	\put(.0,0.25){ \epsfysize=7.0cm
		    \epsffile[100 200 450 600]{null.ps} 
                  }
\end{picture} 
      \caption{(B) Collapse factor ${\cal C}$ 
versus particle number $N$ for uniform spheroidal distributions. 
The collapse factor for axisymmetric spheroids is well fitted with the 
empirical power $N^{1/6}$ (solid line). The $N^{1/3}$-scaling law of spheres 
is also shown for reference, for both uniform- and Plummer- initial
distributions (dotted lines). 
The open triangles are results of Boily et al. (1999) 
obtained with a direct-summation $N$-body algorithm. The horizontal
arrow indicates a collapse factor $\approx O(20)$,  
such that  50\% of all triaxial ellipsoids of infinite particle number
(i.e. pressure-less fluid) collapse to smaller values. 
              }
         \label{fig:scaling2}
   \end{figure*}

Our  computational strategy must therefore ensure that the mass profile is 
well resolved at all times. Because the one-dimensional spherical collapse 
provides the strictest numerical test of our numerical set-up, we first recover the scaling (\ref{eq:scaling}) for large particle number to refine the code's grid resolution. We do this for values of  $N$ ranging from $10^4$ to $16\times 10^6$. 
The mass distribution is mapped accurately by Lagrangian radii sorted 
on concentric shells. 
 We have measured the collapse factor (\ref{eq:factor}) using both the
ratio of gravitational radius, $r_g$, and two  
shells enclosing 30\% and 60\% of the total mass. As can be seen from 
Table~\ref{tab:Spheres}, the collapse factor found for given grid resolution 
and particle number varies considerably depending on the choice of 
Lagrangian radii, as well as with the ratio of gravitational energies. 
 However the free-fall time is recovered to 1.6\% or better. 
These different values obtained for the collapse factor were used to
define error bars on the averaged quantities. The trend with particle
number 
is displayed on Fig.~\ref{fig:scaling}(a). 
Results in spherical symmetry are plotted as 
circles on  the figure: the filled circles  represent results from
this paper, while 
open circles are taken from ALP+88. Each point  is the average of 
data listed in Table~\ref{tab:Spheres} for given $N$, 
but excluding the low-resolution runs.   The scaling 
relation (\ref{eq:scaling}) shown as the solid line on the 
figure provides a good fit to all data points. 

\section{Results for $N$-body spheroidal collapse $(\tau = 0$ or 1)} \label{sec:spheroids} 
Our results for pressure-less fluids show that  initially axisymmetric
 distributions (with $\tau = 0$ or 1) develop spindles  
and divergent gravitational energy as they collapse. These
 configurations  may as a result  show significant
 dependencies on particle number in an $N$-body realisation of the
 solution.  We therefore explore the case of the collapse of 
spheroidal distributions first. 

   \begin{figure*}
\setlength{\unitlength}{2cm} 
\begin{picture}(6,4)(0,0) 
\put(-1.,1.2){ \epsfysize=7.0cm
		    \epsffile[100 200 450 600]{null.ps} 
                  }
\put(3.5,1.2){ \epsfysize=7.0cm
		    \epsffile[100 200 450 600]{null.ps} 
                  }
\end{picture} 
      \caption{Evolution of the self-gravitating energy versus time
         for prolate ($\tau = 0$, left) and oblate ($\tau = 1$,
         right) spheroids. The solid lines give the solution for a
         pressure-less homogeneous fluid; broken lines are $N$-body
         realisations. The labels give the grid cell and particle
         numbers. 
 On the top panels, the set of curves shown at the bottom are the 
dimensionless axes $\tilde{x}_i$ integrated from (\ref{eq:ellip1})
         and(\ref{eq:ellip2}) and we  have indicated the
        times when spindles or discs form. 
 Note how cusps develop in the analytic solution
         each time a disc or spindle forms.} 
         \label{fig:spheroids}
   \end{figure*}

\subsection{Oblate and prolate spheroids} 
To construct spheroidal distributions, we  squeezed or stretched 
the axes 
of spheres to achieve the sought geometry. We consider two cases, 
a prolate $\tau = 0$ spheroid of initial  axes $= (2,1,1)$ and an 
oblate one with axes $= (2,2,1)$. We then performed $N$-body
calculations with $10^4, 10^5$ and $10^6$ particles and compared the 
outcome with the pressure-less fluid solutions for the same initial 
configurations. 

The results are displayed on Fig.~\ref{fig:spheroids}. This graphs the 
gravitational energy $W$ 
as a function of time. The dimensionless axes
$\tilde{x}_i$ of the fluid solution 
are also displayed, 
where we have 
indicated the formation of discs or spindles, according to whether a 
single or two axes vanished at the time indicated. Spindle or disc
each  gives rise to 
sharp  features in the profile of $W$. The numerical 
 solutions for $\tau = 0$ or 1 show clear dependencies on particle
numbers, in the sense that the larger-$N$ calculations map the features 
of the fluid solution more closely. Better agreement 
 with the fluid solutions are to be expected as we 
increase particle number, since the fluid solutions corresponds to
$N\rightarrow \infty$. \mybit{This is difficult to assess quantitatively 
for the solutions as a whole. However we may isolate features that support this 
view.} For instance, as $N$ is increased from $10^4$ to $10^6$
particles, the maximum potential energy achieved in both cases displayed 
increases and corresponds to `spindles' in the fluid solution. 
For the case of the $\tau = 1$ (oblate) fluid spheroid, a spindle forms following 
collapse of the major-axis at $t\simeq 0.623$. The time of maximum potential 
 is $t = 0.646, 0.642 $ and 0.638 respectively for the $N = 10^4, 10^5$ and $10^6$ 
runs, when these maxima shifts upward with $N$, from 9.57 to 13.2, and 15.7 for the largest-$N$ run. The value for the fluid solution $\rightarrow \infty$ formally. Note that the $N = 10^6$ run is the only one with a rapid re-collapse to a disc singularity (see inset at $t\approx 0.64$) similar to the fluid solution. A similar comment can be made for the $\tau = 0$ (prolate) fluid spheroid, where the sequence of singularities is reversed: a spindle forms first, followed by disc and spindle singularities. The two numerical calculations both follow the 
fluid solution relatively well, with one important difference: at the time of the first 
spindle, $t \simeq 0.370$, the $10^4$ particle run shows an increase in $W/W(0)$ much 
reduced compared with the $10^6$ particle run (3.77 to 5.27, or 70\% as much); at $t \simeq 
0.525$ a second spindle forms but now the two $N$-body runs develop very similar extrema in
$W/W(0)$: 5.42 and 5.61, respectively. This means that following rebound through the 
firsts spindle, both calculations suffer a comparable degree of orbit mixing, which then 
 smoothes out the second singularity seen in the fluid. 
 As for the $\tau = 1$ case, the phase of the $N$-body curves tunes up to the 
fluid solution as $N$ increases: thus the first singularity at $t \simeq 0.370$ 
 is found at $t = 0.385\ (N=10^4)$ and $t = 0.376\ (N=10^6)$, which differs with the 
fluid solution by only 1.6\%. 


The two cases $\tau = 0$ and 1 displayed on Fig.~\ref{fig:spheroids} 
   both develop spindles which will reach arbitrarily large 
 potential energy  as $N\rightarrow \infty$. Both  behave 
  in a qualitatively identical way in this respect. 
 We decided therefore to investigate in more
   details only the relation of the solution for the oblate $\tau = 1$ 
case to the number of particles, 
before exploring triaxial initial configurations.

\subsection{Series of oblate spheroids} 
We constructed oblate spheroidal distributions as indicated above. 
The equator of the spheroids lies in the XY plane. 
 The diameter was kept fixed and only one axis resized to achieve the
desired aspect ratio, hence the gravitational energy $W$ is 
magnified with decreasing $a_{3}(0)/a_{1}(0)$. In the limit $a_{3}(0) \ra 0$ we
compute a radial free-fall time 

\[ \tff(a_{3}[0] = 0) = \sqrt{\frac{4}{3\pi}} \times \tff({\rm Eq.}\, [\ref{eq:newtff}]) 
\simeq 0.5116\, . \] 
The time of maximum contraction would therefore lie between this and the
value (\ref{eq:newtff}) for spheres.

\begin{table*}
\centering 
\caption{
Collapse factor  ${\cal C} = W_{max}/W(0)$ 
for uniform spherical distributions. The models have varying
 particle number ($N$), mesh size and linear resolution ($m$ and $l$) in
model units, but the 
same initial total potential energy $W(0) = 1.20$. The free-fall time $\tff = t_{max}$ 
 corresponds to the time when  $W$ reached a maximum; the analytic 
value (\ref{eq:newtff}) is given in round brackets. At that 
time Lagrangian radii enclosing $30\%$ and $60\%$ of the mass were measured; 
their values are given here respective to their initial values, $L_o(30\%) = 0.670$ and $ L_o(60\%) = 0.843$. 
}
\begin{tabular}{crcccccc}
$N$ &  $m$ & $l$ & $\tff = t_{max}$ & $\displaystyle{\frac{W_{max}}{W(0)}}$ & \multicolumn{2}{c}{$L_o/L$}& Comments \\
    &    & $\div 10^{-3}$ & (0.7854)  &         &     $30\%$   & $60\%$   & \\
\hline  \\
$10^4$ & 64 & $1.6$ & 0.795 & 29.5 & 35.1 & 31.9 & \\
$10^4$ & 64 & $1.6$ & 0.799 & 29.6 & 35.1 & 31.9 &
    Reduced $\delta t$ \\
$10^4$ & 32 & $3.1$ & 0.798 & 18.6 & 26.1 & 18.9 & Lower resolution \\ \\
$10^5$ &128 & $0.8$ & 0.787 & 67.0 & 85.9 & 75.0 & \\
$10^5$ & 64 & $1.6$ & 0.788 & 67.6 & 91.8 & 89.2 & \\
$10^5$ & 64 & $1.6$ & 0.788 & 56.1 & 73.8 & 64.4 &$\delta t\times 2$ \\
$10^5$ & 32 & $2.6$ & 0.786 & 32.9 & 58.9 & 41.0 & Lower resolution \\ \\
$10^6$ & 128 & $0.8$ & 0.784 & 97.3 & 133.8 & 112.4 & \\
$10^6$ &  64 & $1.6$ & 0.781 & 73.4 & 121.6 &  89.7 & Lower resolution \\ \\
$1.6\times 10^7$ & 128 & $0.8$ & 0.784 & 225.9 & 352.31 & 290.0 & \\ 
\end{tabular}
\label{tab:Spheres}
\end{table*}
%
%

\begin{table*}
\begin{center} 
\caption{
Collapse factor ${\cal C} = W_{max}/W(0)$ for uniform spheroidal distributions. 
Symbols as for Table~\ref{tab:Spheres}. 
\label{tab:spheroidals} }
\begin{tabular}{crcccccc}
$N$ &  $m$ & $l$ & $t_{max}$ & $\displaystyle\frac{a_{3}(0)}{a_{1}(0)}$ & $\displaystyle{\frac{W_{max}}{W(0)}}$ & Comments \\
    &      &   $\div 10^{-3}$   &        & $< 1 $      &         & \\
\hline  \\
$10^4$ & 32 & $3.1$ & 0.737 & 4:5 & 12.27 & \\
$10^4$ & 32 & $3.1 $ & 0.732 & 4:5 & 11.04 & \\
$10^4$ & 32 & $3.1 $ & 0.697 & 2:3 & 10.23 & \\
$10^4$ & 64 & $3.1 $ & 0.695 & 2:3 & 12.40 & \\
$10^4$ & 32 & $3.1 $ & 0.659 & 1:2 & 11.52 & \\
$10^4$ & 64 & $3.1 $ & 0.650 & 1:2 & 12.65 & \\
$10^4$ & 32 & $3.1 $ & 0.646 & 1:2 &  9.23 & \\
$10^4$ & 32 & $3.1 $ & 0.608 & 1:3 & 11.88 & \\
$10^4$ & 64 & $1.6 $ & 0.602 & 1:3 & 17.26 & \\
$10^4$ & 32 & $3.1 $ & 0.574 & 1:6 &  8.85 & \\
$10^4$ & 32 & $3.1 $ & 0.584 & 1:6 & 10.50 & 
$\leftb W/W_o\rightb = 11.6 \pm 1.3 $ \\
\\
$10^5$ & 32 & $3.1 $ & 0.722 & 4:5 & 17.44 & \\
$10^5$ & 64 & $1.6 $ & 0.724 & 4:5 & 17.86 & \\
$10^5$ &128 & $0.8 $ & 0.563 & 4:5 & 18.09 & \\
$10^5$ & 32 & $3.1 $ & 0.686 & 2:3 & 13.32 & \\
$10^5$ & 32 & $3.1 $ & 0.642 & 1:2 & 13.29 & \\
$10^5$ & 32 & $3.1 $ & 0.597 & 1:3 & 18.40 & \\
$10^5$ & 64 & $3.1 $ & 0.595 & 1:3 & 24.1 & \\
$10^5$ &128 & $1.6 $ & 0.597 & 1:3 & 30.1 & \\
$10^5$ & 32 & $3.1 $ & 0.558 & 1:6 & 14.54 & \\
$10^5$ & 64 & $3.1 $ & 0.559 & 1:6 & 15.69 & \\
$10^5$ & 64 & $1.6 $ & 0.563 & 1:6 & 17.19 & \\
$10^5$ &128 & $0.8 $ & 0.563 & 1:6 & 17.57 & 
$\leftb W/W_o\rightb = 18 \pm 3 $ \\
\\
$10^6$ &128 & $0.8 $ & 0.752 & 9:10 &33.17 & \\
$10^6$ &128 & $0.8 $ & 0.682 & 2:3 & 17.68 & \\
$10^6$ &64 & $0.8 $ & 0.634 & 1:2 & 15.60 & \\
$10^6$ &128 & $0.8 $ & 0.637 & 1:2 & 15.84 & \\
$10^6$ &128 & $0.8 $ & 0.595 & 1:3 & 44.16 & \\
$10^6$ &128 & $0.8 $ & 0.593 & 1:3 & 47.42 & \\
$10^6$ &128 & $0.4 $ & 0.593 & 1:3 & 47.48 & Higher resolution \\
$10^6$ &64  & $1.6 $ & 0.596 & 1:3 & 28.89 & Lower resolution \\
$10^6$ &128 & $0.8 $ & 0.556 & 1:6 & 23.20 &
$\leftb W/W_o\rightb = 28 \pm 11 $ \\
\\
$10^7$ &128 & $0.8 $ & 0.718 & 4:5 & 29.33 & \\ 
$10^7$ &128 & $0.8 $ & 0.680 & 2:3 & 22.55 & \\
$10^7$ &128 & $0.8 $ & 0.595 & 1:3 & 57.45 & \\
$10^7$ &128 & $0.8 $ & 0.553 & 1:6 & 36.92 & $\leftb W/W_o\rightb = 37 \pm 10 $
\end{tabular}
\end{center} 

\end{table*}

The results are listed in Table~\ref{tab:spheroidals}. The errors on 
the collapse factor ${\cal C} = W/W(0)$ are computed from variations
about the mean value for fixed number of particles. 
 The aspect ratios initially lie between 1/6 and 9/10, for particle
numbers 
ranging from $10^4$ to $ 10^7$. All results are graphed as  
triangles on Fig.~\ref{fig:scaling2}(b). (The effect of 
different initial aspect ratios on ${\cal C}$ is discussed 
in \S 5.3.)  We have added points 
obtained from simulations with direct-summation codes by Boily et
al. (1999) for small-$N$ systems to those of the present study (filled
 triangles on the figure). 
 For large-$N$ calculations ($N
 \gtabout 10^5$ and beyond) the black triangles mark a gently
 increasing trend, well matched  with a power-law dependence 
${\cal C} \propto N^{\alpha}$ with $\alpha \approx 1/6$. For $N
\approx 10^4$ or smaller, the fit remains good but note the large 
scatter for points obtained with a direct-summation code (open
triangles on the figure). \mybit{The range of collapse factors 
measured for the 10,000 particle runs
listed in Table~\ref{tab:spheroidals} allows for a multiplicative factor of 
3/2 between maximum and minimum values of $W/W(0)$. The results for the 
direct-summation runs would allow a somewhat larger range, of perhaps 5/2. 
 Whether this is cause for concern is debatable because of the small number 
of runs in this bin; it may be that particle-particle interactions, better 
resolved in the direct-summation scheme, increase the scatter somewhat, though 
not the mean values, which we recover well with the FFT scheme.}

The trend with particle number $N$ for spheroidal distributions 
 is never well fitted with the scaling $\propto N^{1/3}$ of spherical 
distributions, though the data differ by only small factors for
 small-$N$ systems. The results of
 Section~\ref{sec:pressurelessellipsoids} suggest that any increasing 
trend of collapse factor ${\cal C}$ 
 with particle number would be a sensitive function of the symmetry 
of the initial distribution, or of evolution towards axisymmetry in 
the course of evolution. 

\subsection{Another look at the LMS flow} 
We investigated the role played by the Lin-Mestel-Shu instability in
numerical $N$-body calculations of violent relaxation. 
 For homogenous systems, the LMS instability develops  
 as an aspherical system collapses from rest first down its shortest axis. 
  For oblate spheroids, the collapse down the minor axis $z$ occurs in a time 

\begin{equation} t (z = 0 ) \approx \sqrt{\frac{a_{3}(0)}{a_{1}(0)} }
\tff({\rm Eq.} \ref{eq:newtff}) \ ,  \label{eq:tz0} \end{equation} 
 where $\tff$ is the free-fall time for spheres. 
For the initially flattest spheroids in our sample, 
of $a_{3}(0):a_{1}(0) = 1:6$, 
 we compute $t = 0.41\,\tff \approx 0.32$. The data given in
 Table~\ref{tab:spheroidals} and Fig.~\ref{fig:spheroids} \& ~\ref{fig:taus}
  show that 
maximum collapse (or, $W$) occurs at later times, as collapse down the
 major axis sets in. This applies to all 
simulations. Thus self-induced LMS-type of instabilities are 
not sufficient by themselves to halt collapse since the 
 kinetic energy dispersion $\sigma$ grows anisotropically. 
 This anisotropy persists up to the 
time of maximum $W$ and beyond, even for low-$N$ particle number
 calculations, and imprints the virialised equilibrium that 
follows (Boily et al. 1999). 

A question remains which concerns the relative importance of the 
LMS instability compared to the fragmentation modes of instability 
that control collapse in spherical symmetry (cf. Section 2.1). 
 Both types of instability will develop during the collapse of a 
spheroid, however the growth rate of the LMS instability deduced from 
(\ref{eq:tz0}) will be higher for spheroids with initially small 
aspect ratios. In Section 2.2, we have argued that the growth rate 
of the LMS instability is always more rapid than the fragmentation
instability if the initial aspect ratio is less than about 0.75. 
 We would, therefore, expect a signature of this instability 
in the form of a stream at a later stage of collapse, namely the
bounce. We seek out evidence for this in our sample of runs of
Table~\ref{tab:spheroidals}.  
Below we refer to the stream as an `LMS flow', which should 
not be confused with the instability described by Lin and co-workers. 
 \newline

%
   \begin{figure}
\setlength{\unitlength}{1cm} 

\begin{picture}(6,8.)(0,0) 
	\put(1.,0.5){ \epsfysize=6.0cm
		    \epsffile[100 200 450 600]{null.ps} 
                  }
\end{picture} 
      \caption{Normalised collapse factor ${\cal C}$ 
versus initial aspect ratio for all runs. The values ${\cal C}$ taken 
from Table~\ref{tab:spheroidals} were normalised to the mean of their 
respective $N$ series. The open squares are the results for $N =
10^4$; black triangles for all others. The dashed line is the 
 function $\sin\left( \theta + \theta^2 - 2 \right)$, where
$\theta \equiv 2\pi\, \displaystyle{\frac{a_3}{a_1}}$. The histogram at the bottom 
shows the relative phase of the minor-axis and major-axis velocities 
around the time when a singularity forms. 
              }
         \label{fig:eccentricity}
   \end{figure}

Since (\ref{eq:tz0}) shows a relation between minor-axis collapse 
time and 
initial aspect ratio, we expect a similar relation between maximum 
collapse factor ${\cal C} = W_{max}/W(0)$ (the bounce) and initial 
aspect ratio 
for the cases when an LMS flow drives the dynamics at that time. 
 Crucial to the argument is the relative phase of the velocity 
components at the bounce. For the chosen spheroidal initial conditions,  
  the motion can be divided in cylindrical $z$ and $R$ components, and 
we may align the minor axis with the $z$-component of the reference
frame. 
The expectation for LMS flows is that when both $\leftb v_z\rightb$ and
$\leftb v_R\rightb$ are negative inward (in phase) at the bounce, 
the value of $W$  achieved 
should be larger than when $\leftb v_z\rightb$ and $\leftb v_R\rightb$ are of opposite 
signs (out of phase), i.e. one inward, the other outward. 
 Unless fragmentation or other types of instability manage to erase
the signature LMS flow, the relative phase of the velocity components
at the bounce will be set by the initial system aspect ratio. 

We may  compare the ensemble of calculations of Table~\ref{tab:spheroidals}, 
first by normalising individual values of $W/W(0)$ for given $N$ to the 
mean value for that series; in this way we remove the scaling 
$\propto N^{1/6}$ between simulations with different particle
numbers. We then sort the normalised values by increasing
order of the initial aspect ratio, $a_{3}(0)/a_{1}(0)$. 
The results are displayed on Fig.~\ref{fig:eccentricity}. The
sinusoidal pattern of the data is unmistakable. 
 The data may be fitted with an  sinusoid 
 of amplitude $\approx 1/2$, which is comparable or larger than
the intrinsic scatter of the points at given aspect ratio. Thus the 
LMS flow has equal or more impact on the potential and the
system configuration than other factors which predict no dependence
with initial aspect ratio, such as 
the growth of velocity dispersion 
by  internal fragmentation modes (see Section 2). At the bottom of 
Fig.~\ref{fig:eccentricity} we have sketched the relative phase of the 
velocity components obtained for the analytic fluid solution for
oblate spheroids. The step-wise histogram indicates in-phase (high step) 
motion or out-of-phase (low step) motion. The arrows indicate the
polarity of the motion. The correspondence with the numerical 
 data is only suggestive: the pattern itself appears somewhat
out of phase with the histograms. Since the pressure-less fluid
solution from which the histogram was constructed does not suffer from 
any type of instability, the poor agreement with the data would suggest  
 that instabilities other than the LMS instability are not 
completely negligible to set the system properties at the bounce.  

We note that  for $N= 10^4$ particle runs and
initial aspect ratio $> 1/2 $ or so, the data lie near
the normalised ${\cal C} = 1$ value
(cf. Table~\ref{tab:spheroidals} and  Fig.~\ref{fig:eccentricity}, open
squares), hence are not caught in the sinusoidal 
pattern driven by the streaming motion. 
However our data also indicates that $10^4$ particle runs with initial
aspect ratios $< 1/2$ or so match the range of values obtained with 
larger-$N$ runs. 
 Thus fragmentation modes of instability may still be the dominant 
   factor controlling the bounce for simulations with 
particle numbers $\ltabout 10^4$
and initially near-spherical distributions, but not so when the
initial aspect ratio is sufficiently small.

   \begin{figure*}
\setlength{\unitlength}{2cm} 
\begin{picture}(6,8)(0,0) 
	\put(-.3,1.5){ \epsfysize=6.cm
		    \epsffile[100 200 450 600]{null.ps} 
                  }
\put(3.4,2.5){ \begin{minipage}{.4\textwidth} 
\caption{Evolution of the self-gravitating energy $W$ versus time for
three ellipsoidal configurations of initial morphology given by
$\tau$ (eq. [\ref{eq:deftau}]). In all cases, the solid line 
represents the pressure-free fluid solution, while broken lines are
the $N$-body realisations with particle numbers as indicated. The
evolution of the dimensionless ellipsoid axes, $\tilde{x}_i$, are
shown at the bottom for reference. The $N$-body calculations come ever
closer
to the fluid solution with increasing particle number. Note that the 
fluid solution reaches a finite maximum $W$ in all cases displayed. }
         \label{fig:taus}
\end{minipage}                }
\put(-.3,5.2){ \epsfysize=6.0cm
		    \epsffile[100 200 450 600]{null.ps} 
                  }
\put(3.6,5.2){ \epsfysize=6.0cm 
		    \epsffile[100 200 450 600]{null.ps} 
                  }
\end{picture} 
   \end{figure*}

\section{Results for $N$-body triaxial collapse $(\tau \ne 0$ or 1)}
\label{sec:ellipsoids} 
 We now extend our study to triaxial configurations. We repeated the 
exercise of Section~\ref{sec:spheroids}, but this time varying two
initial aspect ratios as parameters. The set-ups used to obtain
numerical and analytic solutions were as before. 
 The results are illustrated on Fig.~\ref{fig:taus}. 
We considered
three triaxial configurations with $\tau = 1/3, 1/2 $ and 2/3, so
covering both prolate and oblate structures. For these cases the 
 analytic fluid solution did not develop spindles and hence the 
potential energy remained finite for the duration of integration. 
 As for the spheroidal calculations, the  numerical $N$-body 
calculations come ever closer to the analytic solution with increasing 
particle number. For example, for $N = 10^6$ particle runs, the 
potential energy already  comes within 20\% of the fluid
solution at maxima. 
 Higher particle numbers would only bring modest
differences and convergence as $N\rightarrow \infty$ is therefore very slow. 
The $10^4$ particles runs remains approximately $50\%$  
out of step with the fluid values, 
and hence the quantities involved with such low-$N$
calculations of relaxation processes are to be treated with caution  
in applications to galaxy or halo formation problems. 

\section{Discussion and conclusion}
 We have sought to constrain the tidal field developing around 
 galaxies and haloes as they form. To do this we studied the 
growth of gravitational energy during the violent relaxation of
 ellipsoidal bodies. We used both an analytic pressure-less gas 
model and $N$-body numerical integration to set absolute limits 
on calculations of galaxy and halo  
formation involving $N$ point masses. 

We found using the pressure-less gas model 
that close to  
9/10 (86\%) of all ellipsoidal triaxial 
configurations increase their gravitational
 energy by at most a factor 40 (cf. Figs.~\ref{fig:Wmax} and
 \ref{fig:distribution}). We confirmed this with $N$-body calculations
 using up to one million particles.  

We studied axisymmetric and spherical uniform distributions. We
extended the scaling of collapse factor  
with particle number for homogeneous spheres, ${\cal C} \propto N^{1/3}$, to 
$N = 16$ millions with the code {\sc superbox}. We noted that
axisymmetric 
spheroidal distributions also show increasing collapse factor with 
particle number: The run of data points is well fitted by the
power-law ${\cal C} \propto N^{1/6}$, much gentler than for
spherically symmetric distributions (see Fig.~\ref{fig:scaling2}[b]). 
\mybit{The scatter in the data 
for spheroids as such does not allow to fall back on the  ${\cal C} \propto N^{1/3}$ scaling 
for spheres, even when we try matching runs with different initial aspect ratio (cf. Table~\ref{tab:spheroidals}). 
}
We pointed out that the extrema
of binding energy and hence tidal forces met by such  systems depend 
sensitively on the formation of prolate structures (spindles) and 
are much gentler otherwise. 

The growth of velocity dispersion during collapse 
can be attributed both to global 
fragmentation modes and Lin-Mestel-Shu-type of pancaking. 
 We presented evidence to the effect that in calculations
involving more than $N = 10^4$ particles, the Poissonian seeds of 
fragmentation modes leads to growth in 
kinetic energy such that the sum remains  smaller than, though not 
negligible before,  the growth of kinetic energy attributable 
to the surface mode (Lin-Mestel-Shu) of instability. 
For $N \ltabout 10^4$ and initial aspect ratio $\gtabout 1/2$, 
the data suggests that fragmentation modes play 
an equally important r\^ole for up to $N = 10^4$ particles (see Fig.~\ref{fig:eccentricity}). 
For $N > 10^5$, the LMS flow at the bounce sets the system properties 
both in terms of potential depth and velocity components for the full 
range of the initial conditions studied here, where aspect ratios where taken 
in the range from 1/6 to 9/10. For this range of particle number and more, 
the collapse of systems with different initial morphologies can be distinguished without ambiguity on Fig.~5, which gives confidence on that 
the gravatitational infall has been properly resolved. \newline

Taken together, these results imply that the formation of
axially- or spherically-symmetric haloes and galaxies lead to deeper 
potentials during the violent relaxation phase and hence to more 
pronounced tidal fields than for non-symmetric ones. The stronger 
tidal fields would in turn reduce the  rate of survival of  
sub-condensations or satellites orbiting within them. 
 Consequently, we would expect  galactic halo morphology and 
 satellite populations to be correlated, in the sense that 
galaxies with rounder massive haloes would harbour fewer   
satellites, while triaxial haloes would harbour  many more satellites, 
 all other parameters being equal.   
This does  not take into account  the long-term fate of galactic 
 satellites: Tidal forces do not subside after virialisation,  and
eventually will cause the disruption of all galactic satellites
 after a period of time 
 (e.g. Ibata et al. 1994, Klessen \& Kroupa 1998, Bullock et
al. 2001). Thus the above statement refers to the time of formation only, 
once the system has virialised. 

Another direct  consequence of our results  is that 
 spherically symmetric haloes should be 
more centrally concentrated than non-spherical ones in virial equilibrium. 
We may expect this to bear on the kinematics of observed galaxies.   
 \mybit{ 
Unfortunately the current observational constraints on the halo shapes
are not sufficiently precise to allow us to test our prediction. Halo
axial ratios have been measured for hardly over a dozen galaxies. What
is more worrisome, however, is that the different techniques seem to
give systematically different results. Merrifield  (2002) summarises
nicely the situation for disc galaxies. Their halos appear to be
axisymmetric and oblate, with their axes of symmetry coalligned with the
disk axes. The most reliable measurements for the minor to major axis
ratios come from polar rings, but galaxies having such structures may
not be a representative sample of disk galaxies, because they might be
the results of recent mergers. Measurements from the flaring of the
 HI disk give
systematically smaller values than those of polar rings measurements.
An application to our own Galaxy (Olling and Merrifield 2000) shows
that such values may be valid only if the value of the distance from
the sun to the Galactic center and the local Galactic rotation speed
are smaller than what is currently believed. Certainly some progress
is necessary before the measurements attain the precision we need for
testing our prediction. }

Gravitational lensing offers some hope by constraining the distribution of 
 total (dark + visible) gravitational mass inside a 
given volume from  the symmetry of the lensed image, which will not 
 respect the centre of mass of the system if it is not spherically symmetric. 
 For instance, Maller et al. (2000) have applied such a lensing technique 
  to the spiral B1600+434. 
  The deconvolution procedure however does suggest that the shape of 
the halo deduced remains dependent on the choice of halo density 
profile (isothermal or otherwise) and symmetry. It may be that the systematic 
application of such techniques to sufficiently large samples would reveal a 
correlation in the sense that we indicated above. 
\newline 

The results obtained for uniform-density distributions should be
contrasted with results obtained for non-uniform initial 
distributions. 
Theis \& Spurzem (1999) investigated the morphological 
evolution of initially cold Plummer distributions. The collapse 
factors they obtained are given on Fig.~\ref{fig:scaling}(a) and
\ref{fig:scaling2}(b). These confirm earlier results by ALP+88 of lower 
values of ${\cal C}$ for non-uniform systems and our own arguments of 
Section~2. The  curve fitting the Theis \& Spurzem data is shown on
Fig.~\ref{fig:scaling2}(b) shifted down \wrt the one obtained for uniform
spheres. This new curve now intersects with the collapse factors
obtained for aspherical distributions and $N \sim 10^3 $
particles. A Plummer model shows an extended envelope of mass density 
 $\rho \propto r^{-5/2}$. Thus for the same particle number, the
collapse factor of a Plummer model is reduced in comparison with a
uniform sphere, presumably due to shell crossing taking place near the
centre. However, Plummer models with larger particle numbers also
collapse by larger factors,  and hence other systems with initially 
steep profiles will, too. In spherical symmetry, the collapse of mass distribution 
with radial dependence (e.g. $\rho \propto r^{-\alpha}$) has bearing on  accretion 
problems, since the mass shells reach the centre a various rates in time. The 
 currently favoured road to galaxy formation would have many clumps converging to 
the centre of mass. Since the scaling we have obtained for uniform-density profiles 
may be extended to non-uniform profiles, as shown with the Plummer model, we may 
hope that the relation of gravitational gradient to initial morphology will 
 also find application to cosmological models of galaxy and galactic halo formation. \newline 

To 
recover the physics of collapsing systems adequately in $N$-body calculations 
requires a
 sufficiently large number of particles in order
 to disentangle effects of mass
distribution and morphology. The results of Fig.~\ref{fig:scaling2}(b) 
suggest a fiducial number $N \sim 100,000$ particles as a clean demarcation 
 (where a large gap appears between the lower dotted and solid line on the
figure) between initially spherical, axisymmetric and triaxial distributions. 
Furthermore, the effect of varying the initial density profile appears only 
to shift the zero-point of the curves, and hence does not affect the relation 
of maximum collapse factor ${\cal C} = W_{max}/W(0)$ to particle number. 

\section*{Acknowledgements} Thanks are due to Albert Bosma and 
 Rachel Somerville for comments on 
a draft version of this paper; and to Albert Bosma for his hindsight on dark matter haloes. 
We are grateful to an anonymous referee 
 for a detailed and constructive report. CMB was funded by the  
      Sonder For\-schungsbereich (SFB)\/ 439 programme in Heidelberg.
An E.G.I.D.E. grant awarded to CMB in 2001 by the French Minist\`ere
des Affaires \'Etrang\`eres helped nbring this project to  completion. 

%

%

\end{document}